\documentclass[prl,twocolumn,superscriptaddress]{revtex4-2}

\usepackage[table,dvipsnames]{xcolor}
\usepackage[colorlinks,linktocpage]{hyperref}
\hypersetup{
	citecolor  = NavyBlue,
	linkcolor  = NavyBlue,
	urlcolor = NavyBlue
}

\usepackage{amssymb,amsmath,latexsym,
multirow,comment,appendix,slashed,color,afterpage,multirow,makecell}
\usepackage{bold-extra}

\pdfoutput=1
\usepackage{bm}
\usepackage{setspace}
\usepackage{tikz}
\usetikzlibrary{calc}

\usepackage{cleveref}
\crefname{table}{Table}{Tables}
\crefname{equation}{Eq.}{Eqs.}
\crefname{appendix}{App.}{Apps.}
\crefname{section}{Sec.}{Secs.}
\crefname{figure}{Fig.}{Figs.}
\usepackage[normalem]{ulem}
\usepackage[most]{tcolorbox}

\makeatletter
\g@addto@macro\bfseries{\boldmath}
\makeatother

\newcommand{\Dcal}{\mathcal{D}}

\newcommand{\dd}{\text{d}}
\newcommand{\pd}{\partial}
\newcommand{\tphi}{\tilde\phi}

\newcommand{\Lag}{\mathcal{L}}

\newcommand{\Sbar}{\overline{S}}
\newcommand{\gbar}{\overline{g}}
\newcommand{\Rbar}{\overline{R}}

\newcommand{\G}{\Gamma}
\newcommand{\Gbar}{\overline{\G}}
\newcommand{\Mcal}{\mathcal{M}}
\newcommand{\Mbar}{\overline{\Mcal}}
\newcommand{\Amp}{\mathcal{A}}
\newcommand{\V}{\mathcal{V}}
\newcommand{\Vbar}{\overline{\V}}
\newcommand{\prop}{\Delta}
\newcommand{\propbar}{\overline{\prop}}
\newcommand{\Prop}{\Xi}

\newcommand{\Mfun}{\mathcal{F}}

\def\eg{\textit{e.g.}}
\def\ie{\textit{i.e.}}
\def\etc{\textit{etc. }}

\definecolor{colorTC}{rgb}{.2,.7,.2}

\newcommand{\s}{\hspace{0.8pt}}

\begin{document}

\begin{flushright}
CERN-TH-2024-176
\end{flushright}
\vspace{1pt}
\title{What is the Geometry of Effective Field Theories?}

\author{Timothy~Cohen}
\affiliation{
Theoretical Physics Department, CERN, 1211 Geneva, Switzerland
}
\affiliation{
Theoretical Particle Physics Laboratory, EPFL, 1015 Lausanne, Switzerland
}
\affiliation{
Institute for Fundamental Science, University of Oregon, Eugene, OR 97403, USA
}

\author{Xiaochuan Lu}
\affiliation{
Department of Physics, University of California, San Diego, La Jolla, CA 92093, USA
}

\author{Zhengkang Zhang\s\s}
\affiliation{
Department of Physics and Astronomy, University of Utah, Salt Lake City, UT 84112, USA
}

\begin{abstract}
\noindent
We elaborate on a recently proposed geometric framework for scalar effective field theories. Starting from the action, a metric can be identified that enables the construction of geometric quantities on the associated functional manifold. These objects transform covariantly under general field redefinitions that relate different operator bases, including those involving derivatives. We present a novel geometric formula for the amplitudes of the theory, where the vertices in Feynman diagrams are replaced by their geometrized counterparts.  This makes the on-shell covariance of amplitudes manifest, providing the link between functional geometry and effective field theories.
\end{abstract}

\maketitle


\begin{center}
{\large\bfseries\scshape Introduction}
\end{center}
\noindent
One of the fundamental difficulties in the Lagrangian formulation of effective field theories (EFTs) is that one can redefine the fields in a theory without changing the physical predictions, \eg\ scattering amplitudes. The freedom to perform field redefinitions to change the form of the Lagrangian can often obscure the physical content of an EFT. This is in fact a familiar situation in any physical theory, where one can pick different coordinate systems for the dynamical degrees of freedom. In the case of classical Hamiltonian mechanics, this corresponds to choosing different coordinates for the same symplectic geometry. What is the analogous geometric picture for EFTs? Is it possible to characterize and classify EFTs by the geometric properties of their associated manifolds? In this letter, we make significant progress towards answering these ambitious questions. 

The notion that fields in an EFT exist on a field manifold has a long history~\cite{Coleman:1969sm, Callan:1969sn, Honerkamp:1971sh, Volkov:1973vd, Tataru:1975ys, Alvarez-Gaume:1981exa, Alvarez-Gaume:1981exv, Vilkovisky:1984st, DeWitt:1984sjp, Gaillard:1985uh, DeWitt:1985sg, Georgi:1991ch}.  This idea, which is often called `field space geometry,' has recently experienced a renaissance, as many new exciting applications have been developed, \eg\ to the Higgs sector of the Standard Model~\cite{Alonso:2015fsp, Alonso:2016btr, Alonso:2016oah, Nagai:2019tgi, Helset:2020yio, Cohen:2020xca, Cohen:2021ucp, Alonso:2021rac, Banta:2021dek, Talbert:2022unj, Alonso:2023jsi, Alonso:2023upf}, EFTs with fermions~\cite{Finn:2020nvn, Pilaftsis:2022las, Gattus:2023gep}, gauge bosons and higher-spin fields~\cite{Finn:2019aip, Helset:2022tlf}, geometric soft theorems~\cite{Cheung:2021yog, Derda:2024jvo}, geometry-kinematics duality~\cite{Cheung:2022vnd}, and renormalization group evolution equations~\cite{Alonso:2022ffe, Helset:2022pde, Assi:2023zid, Jenkins:2023rtg, Jenkins:2023bls}.

One of the primary goals of the field space geometry program is to express the amplitudes in terms of geometric quantities defined on the field manifold.  In this language, field redefinitions are recast as coordinate transformations, and the invariance of on-shell amplitudes can be made manifest by writing them explicitly in terms of geometric quantities.  This approach has led to many insights for understanding the theory and phenomenology of EFTs (\eg\ soft limits of amplitudes~\cite{Cheung:2021yog, Derda:2024jvo} and SMEFT vs.\ HEFT classification of Standard Model extensions~\cite{Alonso:2015fsp, Alonso:2016oah, Cohen:2020xca}).  However, it has a significant limitation in that it only accommodates field redefinitions without derivatives:
\begin{equation}
\phi = f \big( \tphi \big) \,,
\label{eqn:RedefNoD}
\end{equation}
where $f$ is a real analytic function that relates the fields $\phi$ to the fields in the transformed basis $\tphi$, and we are suppressing the flavor index $i$ of the scalar fields $\phi^i$.  On the other hand, it is known that amplitudes are invariant under a much broader class of redefinitions~\cite{Borchers1960, Chisholm:1961tha, Kamefuchi:1961sb, tHooft:1973wag, Arzt:1993gz, Epstein2008, Manohar:2018aog, Criado:2018sdb, Criado:2024mpx}:
\begin{equation}
\phi = f \big( \tphi \;,\; \pd_\mu \tphi \;,\; \pd_\mu \pd_\nu \tphi \;,\; \cdots \big)
= F \big[ \tphi \big] \,,
\label{eqn:Redef}
\end{equation}
where $F$ is a functional of $\tphi$. In other words, the coordinate transformation in \cref{eqn:RedefNoD} must be upgraded to a functional transformation. The set of field redefinitions in \cref{eqn:Redef} are the origin of operator redundancies in EFTs. In this letter, we elaborate on a recently proposed geometry~\cite{Cheung:2022vnd} that accommodates these more generally allowed field redefinitions \footnote{There is also a complementary approach to accommodate derivative field redefinitions utilizing Lagrange geometry and jet bundles~\cite{Craig:2023wni, Craig:2023hhp, Alminawi:2023qtf}.}. We refer to this framework as `functional geometry.'

Some initial explorations towards understanding amplitudes on the functional manifold were recently performed in Refs.~\cite{Cohen:2022uuw, Cohen:2023ekv}, where two of us with collaborators showed that the off-shell amplitudes that have been stripped of their external wavefunction factors (see \cref{eqn:Recursion} below) transform covariantly up to a set of terms that vanish at the physical vacuum with the on-shell conditions enforced.  This `on-shell covariance' of off-shell amplitudes must be accommodated by any geometric picture that describes a non-trivial manifold away from the point that corresponds to the physical vacuum. We emphasize that performing a field redefinition is inherently an off-shell operation, so understanding the true underlying geometry of EFTs requires exploring how off-shell amplitudes transform.

In this work, we build on the metric and the associated geometry introduced in Ref.~\cite{Cheung:2022vnd} in the context of geometry-kinematics duality to elucidate the covariant properties of EFT amplitudes in functional geometry. Starting from this metric, one can define covariant quantities such as curvature tensors, and covariant derivatives thereof. However, since the amplitudes are only on-shell covariant \cite{Cohen:2022uuw, Cohen:2023ekv}, it must be the case that they are not constructed entirely from these tensors. A main result in this letter is that off-shell amplitudes can be constructed from the standard Feynman rules using a geometrized version of the vertices. These new vertices are not tensors on the functional manifold, but are on-shell covariant. Expressing off-shell amplitudes in terms of these novel building blocks makes their on-shell covariance manifest. As a corollary, we will show that our construction reproduces the geometry-kinematics duality~\cite{Cheung:2022vnd} in the massless limit, and leads to a generalization of this duality to massive theories.

The introduction of functional geometry sets the stage to significantly impact our general understanding of EFTs.  By reframing results based on field space geometry in terms of functional geometry, we can explore the robustness of conclusions drawn when using the more limited class of field redefinitions expressed by \cref{eqn:RedefNoD}.  There are many applications to the theory and phenomenology of EFTs, some of which we highlight in the Outlook section below.

\vspace{5pt}
\begin{center}
{\large\bfseries\scshape Geometry of the Functional Manifold}
\end{center}
\vspace{5pt}
\noindent
Field redefinitions of the form \cref{eqn:Redef} correspond to coordinate changes on the functional manifold, $\phi = F[\tphi]$. A key property of these field redefinitions is that they commute with spacetime translations,
\begin{equation}
\mathcal{T}_\epsilon F[\phi] = F[\mathcal{T}_\epsilon\phi] \,,
\label{eqn:TranslationInvariance}
\end{equation}
where $\mathcal{T}_\epsilon\phi(x) = \phi(x+\epsilon) = \phi(x) +\epsilon^\mu \partial_\mu\phi(x) +\mathcal{O}(\epsilon^2)$. As a result, we have
\begin{equation}
\pd_\mu\phi^i(x) = \int \dd^4y\, \frac{\delta\phi^i(x)}{\delta\tphi^j(y)}\, \pd_\mu\tphi^j(y) \,,
\label{eqn:TransChainRule}
\end{equation}
where $i,j$ are flavor indices, and `$\delta$' denotes a functional derivative in the usual sense. \cref{eqn:TransChainRule} shows that $\pd_\mu\phi^i$ transforms as a vector under field redefinitions that can include derivatives.

It is more convenient to use momentum space fields $\phi^i(p) = \int \dd^4x\, e^{ip\cdot x}\, \phi^i(x)$ to chart the functional manifold. Field redefinitions in \cref{eqn:Redef} can then be written as (suppressing flavor indices):
\begin{align}
\phi(p) &= c_0(p) + \int \frac{\dd^4 p_1}{(2\pi)^4}\, c_1(p;p_1)\, \tphi(p_1)
\notag\\[5pt]
&\quad
+ \frac12 \int \frac{\dd^4 p_1}{(2\pi)^4} \frac{\dd^4 p_2}{(2\pi)^4}\, c_2(p;p_1,p_2)\, \tphi(p_1) \tphi(p_2) + \cdots \,,
\label{eqn:RedefMomentumSpace}
\end{align}
with
\begin{equation}
c_n(p; p_1, \cdots, p_n) \propto (2\pi)^4\, \delta^4 
\Bigl(p-\textstyle\sum\limits_{a=1}^n p_a\Bigr) \,.
\label{eqn:cnTransInv}
\end{equation}
From \cref{eqn:RedefMomentumSpace,eqn:cnTransInv}, we find the momentum space version of \cref{eqn:TransChainRule}:
\begin{equation}
p_\mu \phi^i(p) = \int \frac{\dd^4 q}{(2\pi)^4}\, \frac{\delta\phi^i(p)}{\delta\tphi^j(q)}\, q_\mu \tphi^j(q) \,,
\label{eqn:TransChainRulep}
\end{equation}
which implies that $p_\mu \phi^i(p)$ transforms as a vector. We emphasize that the vector transformation property in \cref{eqn:TransChainRule,eqn:TransChainRulep} is a consequence of translation invariance of the field redefinitions in \cref{eqn:Redef}.

The EFT action $S[\phi]$ is a scalar on the functional manifold. Therefore, if we can write~\cite{Cheung:2022vnd}
\begin{equation}
S = -\frac12 \int \frac{\dd^4 p}{(2\pi)^4} \frac{\dd^4 q}{(2\pi)^4}\, g_{ij}(p,q)\, p_\mu\phi^i(p)\, q^\mu \phi^j(q) \,,
\label{eqn:Metric}
\end{equation}
with $g_{ij}(p,q) = g_{ji}(q,p)$, then $g_{ij}(p,q)$ can be identified as a metric on the functional manifold, and as such it transforms as a $(0,2)$-tensor, \ie\ $\tilde g_{ij}(p,q) = \int \frac{\dd^4 r}{(2\pi)^4} \frac{\dd^4 s}{(2\pi)^4}\,\frac{\delta\phi^k(r)}{\delta\tilde\phi^i(p)}\,\frac{\delta\phi^l(s)}{\delta\tilde\phi^j(q)}\, g_{kl}(r,s)$. For example, $g_{ij}(p,q) = (2\pi)^4\delta^4(p+q) \,\delta_{ij} (1-m_i^2/p^2)$ corresponds to a theory of free scalars. Note that for a given action, \cref{eqn:Metric} does not uniquely determine the metric~\cite{Cheung:2022vnd}; two viable metrics can differ by a tensor $h_{ij}(p,q)$ that satisfies $\int \frac{\dd^4 p}{(2\pi)^4} \frac{\dd^4 q}{(2\pi)^4}\,h_{ij}(p,q)\, p_\mu\phi^i(p) \,q^\mu\phi^j(q) = 0$. We will come back to discuss this ambiguity in the Outlook section.

Starting from a metric, we can construct the Christoffel connection, covariant derivative, and Riemann curvature tensor in the usual way. In what follows, we abbreviate $\phi^{i_a}(p_a) \equiv \phi^a$, $g_{i_1i_2} (p_1, p_2) \equiv g_{12}^{}$, and similarly for all the other objects. In this notation, derivatives on the functional manifold are normalized as
\begin{equation}
{}^{}_{,a} \equiv (2\pi)^4 \frac{\delta}{\delta\phi^a} \,,
\end{equation}
and index contraction means
\begin{equation}
{}^b{}_b \equiv \sum_{i_b} \int \frac{\dd^4 p^{}_b}{(2\pi)^4} \,.
\end{equation}
The Christoffel connection is given by
\begin{equation}
\G^a_{12} = \frac12\, g^{ab} (g^{}_{b1,2} + g^{}_{b2,1} - g^{}_{12,b}) \,,
\end{equation}
where the inverse metric $g^{ab}$ is defined from $g^{ab} g^{}_{bc} = \delta_a^c \equiv \delta_{i_a}^{i_c} (2\pi)^4 \delta^4(p_a-p_c)$, and the Riemann curvature tensor is
\begin{equation}
R_{1234} =\frac12\, (g_{14,23} - g_{24,13}) - g_{ab}\G^a_{13} \G^b_{24} -(3\leftrightarrow 4) \,.
\end{equation}

We are often interested in the values of geometric quantities at the special point on the functional manifold corresponding to the physical vacuum $\bar\phi$. Using overline to denote quantities evaluated at $\bar\phi$, we have $\Sbar_{;a} = \Sbar_{,a} = 0$ at tree level, where semicolon denotes the covariant derivative. Without loss of generality, we focus on coordinate choices where $\bar\phi = 0$. From the definition of the metric in \cref{eqn:Metric}, one can readily show:
\begin{subequations}\label{eqn:SandR}
\begin{align}
\Sbar_{;123} &= 0 \,, \\[5pt]
\Sbar_{;(1234)} &= -\frac23\s \Big[ s_{12} \Rbar_{1(34)2} + s_{13} \Rbar_{1(24)3} + s_{14} \Rbar_{1(23)4} \Big] \,, \notag\\[2pt]
\end{align}
\end{subequations}
where $s_{ab}\equiv(p_a+p_b)^2$ are Mandelstam variables and parentheses denote symmetrization of the indices, \eg\ $\Rbar_{1(34)2} = \frac{1}{2} \bigl(\Rbar_{1342} + \Rbar_{1432}\bigr)$.

\vspace{10pt}
\begin{center}
{\large\bfseries\scshape Geometrizing Amplitudes}
\end{center}
\vspace{5pt}
\noindent
Given an EFT action, one can recursively construct the off-shell amplitudes $\Mcal_{1\cdots n}$~\footnote{$\Mcal_{1\cdots n}$ is the Fourier transform of the quantity defined in Eq. (2.22) of Ref.~\cite{Cohen:2023ekv}, up to an overall minus sign.}
using standard Feynman rules \cite{Cohen:2022uuw, Cohen:2023ekv}:
\begin{subequations}\label{eqn:Recursion}
\begin{align}
\Mcal_{123} &= V_{123} \,, \\[5pt]
\Mcal_{1\cdots n (n+1)} &= \Mcal_{1\cdots n, n+1} - \sum_{k=1}^n \prop^{ab}\, V_{b(n+1)k}\, \Mcal_{1\cdots \slashed{k} a \cdots n} \,.
\end{align}
\end{subequations}
We use `$\slashed{k}$' to denote the absence of index $k$. In these equations, $V_{1\cdots k}$ are one-particle-irreducible (1PI) vertices and $\prop^{ab}$ is the full propagator. At tree level,
\begin{equation}\label{eqn:V_Delta}
V_{1\cdots k} = S_{,1\cdots k}
\,,\qquad\text{and}\qquad
\prop^{ab} S_{,bc} = \delta_c^a \,.
\end{equation}
At loop level, one replaces the classical action $S$ in \cref{eqn:V_Delta} by the 1PI effective action. From \cref{eqn:Recursion}, we can write down the $n$-point off-shell amplitude as a function of propagators $\Delta^{ab}$ and vertices $V_{a_1\cdots a_k}$ with $k\le n$:
\begin{equation}
\Mcal_{1\cdots n} = \Mfun_{1\cdots n}\, \Big( \prop^{ab} \;,\; \big\{ V_{a_1\cdots a_k} \big\} \Big) \,.
\label{eqn:MinV}
\end{equation}
For example,
\begin{subequations}\label{eqn:M3M4inV}
\begin{align}
\Mcal_{123} &= V_{123} \,, \\[5pt]
\Mcal_{1234} &= V_{1234} - \prop^{ab} \big( V_{b41} V_{a23} + V_{b42} V_{1a3} 
+ V_{b43} V_{12a} \big) \,. \notag\\[2pt]
\end{align}
\end{subequations}
On-shell amplitudes are then obtained via the Lehmann-Symanzik-Zimmermann (LSZ) reduction formula \cite{Lehmann:1954rq, Lehmann:1957zz} by evaluating $\Mcal_{1\cdots n}$ at the physical vacuum $\bar\phi$ for on-shell external momenta $p_i^2 = m_i^2$, and dressing it with wavefunction factors:
\begin{align}
(2\pi)^4 \delta^4(p_1+\cdots+p_n)\, &\Amp_{i_1 \cdots i_n}(p_1, \cdots, p_n) \notag\\[5pt]
&\hspace{-40pt}
= \Big( r_1^{1/2} \cdots r_n^{1/2} \Big)\, \Big( \Mbar_{1 \cdots n} \big|_\text{on-shell} \Big) \,.
\label{eqn:AinM}
\end{align}

Under generic field redefinitions, the off-shell amplitudes do not transform covariantly:
\begin{equation}
\widetilde{\Mcal}_{a_1\cdots a_n} = \bigg( \frac{\delta\phi^{b_1}}{\delta\tphi^{a_1}} \cdots \frac{\delta\phi^{b_n}}{\delta\tphi^{a_n}} \bigg)\, \Mcal_{b_1\cdots b_n}
+ X_{a_1\cdots a_n} \,,
\label{eqn:McalTrans}
\end{equation}
with $X_{a_1\cdots a_n} \ne 0$. However, one can show recursively that~\cite{Cohen:2023ekv}:
\begin{equation}
\overline{X}_{a_1\cdots a_n} \big|_\text{on-shell} = 0 \,.
\label{eqn:Xvanishes}
\end{equation}
In other words, $\Mcal_{1\cdots n}$ are `on-shell covariant' \footnote{Objects like $X_{a_1\cdots a_n}$ that vanish on-shell at the physical vacuum are referred to as `evanescent' in Ref.~\cite{Cohen:2023ekv}.}.
Meanwhile, the $r_i^{1/2}$ factors in \cref{eqn:AinM} transform as vielbeins~\cite{Cheung:2021yog, Cohen:2023ekv}, so on-shell amplitudes $\Amp_{i_1 \cdots i_n}(p_1, \cdots, p_n)$ for any given 
set of particle species and momenta are invariant under field redefinitions.

The main new result of this work is that, starting from \cref{eqn:Recursion}, it can be shown recursively (see Appendix) that $\Mcal_{1\cdots n}$ can in fact be constructed from a set of on-shell-covariant building blocks:
\begin{equation}
\hspace{15pt}
    \begin{minipage}{0.35\textwidth}
    \begin{center}
        \begin{tcolorbox}[colback=white]
        \vspace{-13pt}
            \begin{equation*}
            \Mcal_{1\cdots n} = \Mfun_{1\cdots n} \Big( \prop^{ab} , \big\{ \V_{a_1\cdots a_k} \big\} \Big) \,.
            \label{eqn:MinVcal}
            \end{equation*}
        \vspace{-18pt}
        \end{tcolorbox}
    \end{center}
    \end{minipage}
\end{equation}
Here $\Mfun_{1\cdots n}$ is the same function as in \cref{eqn:MinV}, but the arguments $V_{a_1\cdots a_k}$ are replaced by a new set of vertices $\V_{a_1\cdots a_k}$. At tree level, these new vertices are given by
\begin{equation}
\V_{1\cdots k} \equiv S_{;1\cdots k} + \G_{1\cdots k}^a S_{,a} + \sum_{b\in\text{external}} \G_{1\cdots \slashed{b} \cdots k}^a S_{,ab} \,,
\label{eqn:Vcaldef}
\end{equation}
where the sum is over the subset of $\{1,\cdots, k\}$ that correspond to external legs, and $\G_{1\cdots k}^a$ are the generalized Christoffel symbols defined recursively via
\begin{equation}
\G^a_{1\cdots k(k+1)} \equiv \G^a_{1\cdots k,k+1} - \sum_{b=1}^k \G^c_{b(k+1)} \G^a_{1\cdots\slashed{b} c \cdots k} \,.
\end{equation}
At loop level, one again replaces the action $S$ in \cref{eqn:Vcaldef} by the 1PI effective action. Note from \cref{eqn:Vcaldef} that when the Christoffel connection vanishes, $\V_{1\cdots k}$ reduce to $V_{1\cdots k}$. Therefore, $\V_{1\cdots k}$ are a geometric version of the standard Feynman vertices $V_{1\cdots k}$, and \cref{eqn:MinVcal} can be viewed as a geometrization of \cref{eqn:MinV}. Importantly, the building blocks in \cref{eqn:MinVcal}, $\prop^{ab}$ and $\V_{1\cdots k}$, are all on-shell covariant; on-shell covariance of $\prop^{ab}$ follows from $\Sbar_{,ab} = \Sbar_{;ab}$, while the on-shell covariance of $\V_{1\cdots k}$ can be proved by induction, which we detail in a follow-up paper~\cite{Paper2}. Since sums and products of on-shell-covariant objects are also on-shell covariant, \cref{eqn:MinVcal} expresses EFT amplitudes in a manifestly on-shell covariant form!

In the following sections, we specialize to the cases of massless and massive theories, respectively. A simple example to illustrate the application of our formalism is included in the Appendix.

\vspace{5pt}
\begin{center}
{\large\bfseries\scshape Massless Theories}
\end{center}
\vspace{5pt}
\noindent
In massless theories, $\Gbar{}_{1\cdots \slashed{b} \cdots k}^a \Sbar_{,ab} = p_b^2\, \gbar_{ab}\, \Gbar{}_{1\cdots \slashed{b} \cdots k}^a$ vanishes as the external momentum $p_b$ goes on-shell, as long as $\gbar_{ab}\, \Gbar{}^a_{1\cdots \slashed{b}\cdots k} \big|_\text{on-shell}$ is finite. This implies
\begin{equation}
\Vbar_{1\cdots k}\, \bigr|_\text{on-shell} = \Sbar_{; 1\cdots k}\, \bigr|_\text{on-shell}
\qquad\text{(massless)} \,.
\label{eqn:VbarTensor}
\end{equation}
The only possible exception is the $k=3$ case with $1,2,3$ all external, because $\gbar_{a1}\, \Gbar{}^a_{23}$ (and permutations) may diverge on-shell due to the special 3-point kinematics. Consequently,
\begin{align}
\Mbar_{1\cdots n}\, \bigr|_\text{on-shell} &= 
\Mfun_{1\cdots n} \big( \overline{\Prop}{}^{ab} , \big\{ \Sbar_{;a_1\cdots a_k} \big\} \big)\bigr|_\text{on-shell}
\label{eqn:MbarinT}
\end{align}
for all $n\ge 4$, where $\Prop^{ab}$ is the covariant propagator satisfying $\Prop^{ab} S_{;bc} = \delta^a_c$ (the covariant version of $\prop^{ab} S_{,bc} = \delta^a_c$), and we have $\propbar{}^{ab} = \overline{\Prop}{}^{ab}$ since $\Sbar_{,bc} = \Sbar_{;bc}$. \cref{eqn:MbarinT} tells us that in massless theories, $\Mcal_{1\cdots n}$ are actually on-shell equivalent to a set of tensors built from $\Prop^{ab}$ and $\{S_{;a_1\cdots a_k}\}$.

An interesting special case of massless theories is the nonlinear sigma model (NLSM):
\begin{equation}
S_\text{NLSM} \equiv \frac12\int \dd^4x\, \text{\bf g}_{ab}(\phi) (\pd_\mu \phi^a) (\pd^\mu \phi^b) \,,
\end{equation}
where $\text{\bf g}_{ab}$ is the metric in field space geometry. Comparing with \cref{eqn:Metric} we see that
\begin{equation}
g_{ij}(p,q) = \int \dd^4x \, e^{-i(p+q)\cdot x}\,\text{\bf g}_{ij}\bigl(\phi(x)\bigr)\,.
\end{equation}
Taking successive functional derivatives and setting $\phi=\bar\phi$, we obtain
\begin{equation}
\gbar_{ab,1\cdots k} = (2\pi)^4 \delta^4(p_a+p_b+p_1+\cdots p_k)\, \overline{\text{\bf g}}_{i_a i_b, i_1\cdots i_k} \,.
\label{eqn:gbar_relation}
\end{equation}
An immediate consequence of \cref{eqn:gbar_relation} is that in the NLSM, any expression in functional geometry that is a sum of contractions of $\gbar_{ab,1\cdots k}$ with $\gbar^{ab}$ (potentially also multiplied by $p_a^\mu$ factors) will be equal to the corresponding expression in field space geometry in terms of $\,\overline{\text{\bf g}}_{i_a i_b,i_1\cdots i_k}$ and $\,\overline{\text{\bf g}}^{i_a i_b}$, multiplied by an overall momentum-conserving $\delta$-function. Off-shell amplitudes evaluated at the physical vacuum $\overline{\Mcal}_{1\cdots n}$ are such expressions. Importantly, from \cref{eqn:Metric,eqn:MinVcal} we see that a universal set of such expressions in terms of $\gbar_{ab,1\cdots k}$ and $\gbar^{ab}$ will give $\overline{\Mcal}_{1\cdots n}$ for {\it all\s} theories, while \cref{eqn:gbar_relation} implies that the same expressions in terms of $\s\overline{\text{\bf g}}_{i_a i_b,i_1\cdots i_k}$ and $\overline{\text{\bf g}}^{i_a i_b}$ will give the correct $\overline{\Mcal}_{1\cdots n}$ in the NLSM. In other words, to find the universal expressions for $\Mbar_{1\cdots n}$ in terms of $\gbar_{ab,1\cdots k}$ and $\gbar^{ab}$, one can simply take the field space geometry expressions for $\overline{\Mcal}_{1\cdots n}$ in the NLSM, and make the replacement $\overline{\textbf{g}}\to \gbar$. This is the geometry-kinematics duality proposed for massless theories in Ref.~\cite{Cheung:2022vnd}. From our argument above, it is clear that the same strategy of replacing $\overline{\textbf{g}}\to \gbar$ should yield the correct amplitudes in both massless and massive theories. The special feature in the massless case is that 4- and higher-point {\it on-shell} amplitudes can be written entirely in terms of tensors on the functional manifold as we saw above. They can therefore be obtained from the tensorial expressions of on-shell NLSM amplitudes in field space geometry, by upgrading the curvature tensors and their covariant derivatives to their counterparts in functional geometry. See Ref.~\cite{Cheung:2022vnd} for many explicit examples. On the other hand, to obtain 3-point amplitudes in massless theories, as well as amplitudes in general massive theories, one must perform the aforementioned replacement in the non-tensorial expressions of {\it off-shell} NLSM amplitudes.

\vspace{5pt}
\begin{center}
{\large\bfseries\scshape Massive Theories}
\end{center}
\vspace{5pt}
\noindent
For massive theories, $\Gbar{}_{1\cdots \slashed{b} \cdots k}^a \Sbar_{,ab} = p_b^2\, \gbar_{ab}\, \Gbar{}_{1\cdots \slashed{b} \cdots k}^a$ does not vanish as $p_b$ goes on-shell, so 
\begin{equation}
\Vbar_{1\cdots k} \bigr|_\text{on-shell} \ne \Sbar_{; 1\cdots k} \bigr|_\text{on-shell}
\qquad\text{(massive)} \,,
\end{equation}
and $\Mcal_{1\cdots n}$ are generically not on-shell equivalent to the tensors $\Mfun_{1\cdots n} \big( \Prop^{ab} , \{ S_{;a_1\cdots a_k} \} \big)$, \ie, \cref{eqn:MbarinT} does not hold generically. As discussed above, this does not invalidate a geometry-kinematics duality, as long as we implement the replacement $\overline{\textbf{g}}\to \gbar$ in the correct non-tensorial expressions. However, we emphasize that our main result in \cref{eqn:MinVcal} organizes these non-tensorial expressions into on-shell-covariant building blocks, making the on-shell covariance of $\Mcal_{1\cdots n}$ manifest.

Another perspective on the presence of non-tensorial $p_b^2\, \gbar_{ab} \,\Gbar{}^a_{1\cdots \slashed{b}\cdots k}$ terms can be gained by going to Riemann normal coordinates $\eta^a$:
\begin{equation}
\phi^a = \eta^a - \sum_{k=2}^\infty \frac{1}{k!} \, \Gbar{}^a_{(b_1\cdots b_k)}\, \eta^{b_1} \cdots \eta^{b_k}\,.
\label{eqn:RNC}
\end{equation}
In the $\eta^a$ basis, \cref{eqn:VbarTensor,eqn:MbarinT} are satisfied since covariant derivatives are the same as ordinary derivatives in normal coordinates, and there are no non-tensorial terms. However, the nonlocal field redefinition in \cref{eqn:RNC} does not belong to the set of field redefinitions in \cref{eqn:Redef}, so naively applying the LSZ reduction formula \cref{eqn:AinM} in the $\eta^a$ basis would not give the correct on-shell amplitudes. The difference can be systematically calculated by working out modifications to \cref{eqn:AinM} due to the introduction of additional single-particle poles, and we have checked that they exactly reproduce the non-tensorial terms; see Ref.~\cite{Paper2} for details.

We emphasize that the discussion above does not preclude the possibility that $\Mcal_{1\cdots n}$ in massive theories are on-shell equivalent to some other tensors constructed differently from those in massless theories. For example, one might construct an alternative connection $\G'^a_{12}$ which, unlike the metric-compatible connection $\G^a_{12}$, does not have a pole at $p_a^2=m_a^2$. Such a connection would ensure $\Gbar{}'^a_{1\cdots \slashed{b} \cdots k} \Sbar_{,ab}$ vanishes on-shell, thereby resurrecting \cref{eqn:MbarinT} when covariant derivatives are taken under $\G'^a_{12}$. We leave an investigation of this possibility to future work.

\vspace{5pt}
\begin{center}
{\large\bfseries\scshape Outlook}
\end{center}
\vspace{5pt}
\noindent
In this letter, we have elaborated on the construction of `functional geometry' for scalar EFTs. We have shown how to write off-shell amplitudes as functions of on-shell-covariant objects on the functional manifold. These functions (\cref{eqn:MinVcal}) are identical to those constructed using the usual Feynman propagators and vertices (\cref{eqn:MinV}), but with the vertices $V_{a_1\cdots a_k}$ replaced by their geometrization $\V_{a_1\cdots a_k}$ given in \cref{eqn:Vcaldef}. In this way, the on-shell covariance of amplitudes is made manifest. We have also discussed how our formalism leads to a more general version of the geometry-kinematics duality~\cite{Cheung:2022vnd}, which applies to all amplitudes in both massless and massive theories.

The discussion in this letter focused on scalar EFTs. It would be interesting to extend the same approach to accommodate higher-spin fields, which we leave to future work. Another avenue worth further investigation is the ambiguity associated with metric choice: \cref{eqn:Metric} does not uniquely determine $g_{ij}(p,q)$. Interestingly, the values of $R_{abcd;1\cdots k}$ and $S_{;1\cdots k}$, as well as our new geometrized vertices $\V_{1\cdots k}$ generally vary with the choice of the metric. This is even the case for their values at the physical vacuum with on-shell momenta~\footnote{For curvature tensors, this metric choice dependence was already noted in Ref.~\cite{Cheung:2022vnd}.}. The metric choice dependence drops only after these building blocks are assembled into the off-shell amplitudes as in \cref{eqn:MinVcal}. Note that this does not invalidate \cref{eqn:MinVcal} as a manifest proof of the on-shell covariance of $\Mcal_{1\cdots n}$, because for the new Lagrangian obtained by a field redefinition, we know that there exists one choice of metric that would correspond to the tensor transformation of the old one. Sticking to this choice, each building block will be on-shell covariant. On the other hand, the meaning of the geometrized vertices $\V_{1\cdots k}$ is somewhat obscured by this metric choice dependence. We know that in field space geometry, the covariantized vertices correspond to vertices in normal coordinates which are directly related to the contact terms in amplitudes. It would be interesting to see if there is an analogous interpretation of $\V_{1\cdots k}$ in functional geometry.

A set of more ambitious goals are to probe the nature of different EFTs and relations between their amplitudes with the functional geometry framework. Concretely, there are established criteria in the language of field space geometry that characterize if an EFT has global symmetries~\cite{Cheung:2021yog}, is renormalizable or non-renormalizable (\eg\ SM vs.\ SMEFT or HEFT)~\cite{Alonso:2015fsp, Alonso:2016oah}, whether a symmetry is linearly or nonlinearly realized (\eg\ SMEFT vs.\ HEFT)~\cite{Coleman:1969sm, Callan:1969sn, Alonso:2015fsp, Alonso:2016oah, Cohen:2020xca}, \etc\ Upgrading these results with functional geometry will take us closer to establishing rigorous criteria for classifying EFTs. Meanwhile, the fact that amplitudes in all scalar EFTs can be constructed from on-shell-covariant building blocks may be exploited to shed new light on the relations between amplitudes in different EFTs. We have only begun to understand the broad implications of the new perspective that functional geometry brings to the dynamics of EFTs.

\vspace{5pt}
\begin{center}
{\large\bfseries\scshape Acknowledgments}
\end{center}
\vspace{5pt}
\noindent
We thank Dave Sutherland for helpful 
discussions and feedback on a preliminary draft.
We thank Andreas Helset for insightful discussions, especially with regards to the extension of geometry-kinematics duality to massive theories.
We thank Matthew Forslund for discussions on field redefinitions and, in particular, for drawing our attention to Ref.~\cite{tHooft:1973wag}.
We thank Yu-Tse (Alan) Lee for informing us of his interesting concurrent development combining supersymmetry and jet bundles to accommodate derivative field redefinitions~\cite{Lee:2024xxx}.
T.C.\ is supported by the U.S.~Department of Energy Grant No.~DE-SC0011640.
X.L.\ is supported in part by the U.S.~Department of Energy Grant No.DE-SC0009919 and in part by Simons Foundation Award 568420. 
Z.Z.\ is supported in part by the U.S.~National Science Foundation under grant PHY-2412880. 
This work was performed in part at the Aspen Center for Physics, which is supported by National Science Foundation grant PHY-2210452.


\bibliographystyle{utphys}
\bibliography{ref}

\onecolumngrid
\newpage

\begin{center}
{\large\bfseries\scshape Appendix}
\end{center}

\renewcommand{\theequation}{A.\arabic{equation}}
\setcounter{equation}{0}

\subsection{1. Proof of \cref{eqn:MinVcal}}
To prove \cref{eqn:MinVcal} by induction, we first note that it is true for $n=3$:
\begin{equation}
    \Mcal_{123} = V_{123} = \V_{123} \,,
\end{equation}
where $\V_{123}$ is obtained from \cref{eqn:Vcaldef} with $1,2,3$ all external. 
Now suppose $\Mcal_{1\cdots n}$ is given by:
\begin{equation}
     \Mfun_{1\cdots n}\, \Big( \prop^{ab} \;,\; \big\{ V_{a_1\cdots a_k} \big\} \Big) = \Mfun_{1\cdots n}\, \Big( \prop^{ab} \;,\; \big\{ \V_{a_1\cdots a_k} \big\} \Big) \,.
    \label{eqn:Mn}
\end{equation}
To obtain $\Mcal_{1\cdots n(n+1)}$, we define a family of linear differential operators $\Dcal^{(\gamma)}$ which, when acting on an object $T^{a\cdots}{}_{b\cdots}$ on the functional manifold, gives:
\begin{align}
    \Dcal^{(\gamma)}_s T^{a\cdots}{}_{b\cdots} &\equiv T^{a\cdots}{}_{b\cdots,s} +\sum_{c\in\text{internal}} \gamma^c_{sd} \,T^{a\cdots\slashed{c}d\cdots}{}_{b\cdots} -\sum_{d\in\text{internal}} \gamma^c_{sd}\, T^{a\cdots}{}_{b\cdots\slashed{d}c\cdots}  \notag\\[5pt]
    & \qquad+\sum_{c\in\text{external}} \Delta^{ce}\,V_{esd}\,T^{a\cdots\slashed{c}d\cdots}{}_{b\cdots} 
    -\sum_{d\in\text{external}} \Delta^{ce}\,V_{esd}\, T^{a\cdots}{}_{b\cdots\slashed{d}c\cdots} \notag\\[10pt]
    &=  T^{a\cdots}{}_{b\cdots,s} + (\gamma^a_{sd}\, T^{d\cdots}{}_{b\cdots} + \,\cdots \,) 
    -(\gamma^c_{sb}\, T^{a\cdots}{}_{c\cdots} + \,\cdots\,) \notag\\[5pt]
    &\qquad+\sum_{c\in\text{external}} (\Delta^{ce}\,V_{esd} -\gamma^c_{sd}) \,T^{a\cdots\slashed{c}d\cdots}{}_{b\cdots} 
    -\sum_{d\in\text{external}} (\Delta^{ce}\,V_{esd} -\gamma^c_{sd})\, T^{a\cdots}{}_{b\cdots\slashed{d}c\cdots}\,.
\end{align}
In other words, we let $\gamma^c_{sd}$ and $\Delta^{ce} V_{esd}$ play the role of connection for internal and external indices, respectively (although we do not require that they transform as connections).
According to \cref{eqn:Recursion}, $\Dcal^{(\gamma)}_{n+1}\Mcal_{1\cdots n}=\Mcal_{1\cdots n(n+1)}$ independently of $\gamma$ (since $\Mcal_{1\cdots n}$ only carries external indices). 
Taking the expression of $\Mcal_{1\cdots n}$ in terms of the standard vertices $V_{a_1\cdots a_k}$ (left-hand side of \cref{eqn:Mn}) and picking $\gamma^c_{sd} = 0$, we obtain an expression for $\Mcal_{1\cdots n(n+1)}$ in terms of:
\begin{subequations}
\label{eqn:D0}
\begin{align}
    \Dcal^{(0)}_{n+1} \Delta^{ab} &= -\Delta^{ac}\Delta^{bd} V_{cd(n+1)} \,,\\[5pt]
    \Dcal^{(0)}_{n+1} V_{a_1\cdots a_k} &= V_{a_1\cdots a_k(n+1)} - \hspace{-5pt}\sum_{b\in\text{external}}\hspace{-5pt} \Delta^{ca} V_{ab(n+1)} V_{a_1\cdots\slashed{b} c \cdots a_k} \,.
\end{align}
\end{subequations}
On the other hand, taking the expression of $\Mcal_{1\cdots n}$ in terms of the geometrized vertices $\V_{a_1\cdots a_k}$ (right-hand side of \cref{eqn:Mn}) and picking $\gamma^c_{sd} = \Gamma^c_{sd}$ (a connection on the functional manifold), we obtain an expression for $\Mcal_{1\cdots n(n+1)}$ in terms of:
\begin{subequations}
\begin{align}
    \Dcal^{(\Gamma)}_{n+1} \Delta^{ab} &= -\Delta^{ac}\Delta^{bd} \V_{cd(n+1)} \,,\\[5pt]
    \Dcal^{(\Gamma)}_{n+1} \V_{a_1\cdots a_k} &= \V_{a_1\cdots a_k(n+1)} - \hspace{-5pt}\sum_{b\in\text{external}}\hspace{-5pt} \Delta^{ca} \V_{ab(n+1)} \V_{a_1\cdots\slashed{b} c \cdots a_k} \,,
\end{align}
\end{subequations}
which are identical to \cref{eqn:D0} with $V$ replaced by $\V$. 
Therefore, if we apply $\Dcal_{n+1}^{(0)}$ to the left-hand side of \cref{eqn:Mn} and $\Dcal_{n+1}^{(\Gamma)}$ to the right-hand side of \cref{eqn:Mn} (which are equivalent operators when acting on $\Mcal_{1\cdots n}$), we obtain:

\begin{equation}
    \Mfun_{1\cdots n(n+1)}\, \Big( \prop^{ab} \;,\; \big\{ V_{a_1\cdots a_k} \big\} \Big) = \Mfun_{1\cdots n(n+1)}\, \Big( \prop^{ab} \;,\; \big\{ \V_{a_1\cdots a_k} \big\} \Big) \,,
\end{equation}
completing the proof.

\newpage

\subsection{2. Geometry of the $\phi^3+\phi^4$ theory}
\noindent
In this appendix we apply the general formalism discussed in the main text to a concrete example. We take the renormalizable Lagrangian of a single flavor scalar field $\phi$:
\begin{equation}
\Lag = \frac12\, (\pd_\mu \phi) (\pd^\mu \phi) - \frac12\, m^2 \phi^2 - \frac16\, \mu\, \phi^3 - \frac{1}{24}\, \lambda\, \phi^4 \,.
\end{equation}
The corresponding action is
\begin{align}
S &= - \int \frac{\dd^4 q_1}{(2\pi)^4} \frac{\dd^4 q_2}{(2\pi)^4}\, (2\pi)^4 \delta^4(q_{12})\, \phi(q_1) \phi(q_2)\, \frac12 \bigl(q_1 q_2 + m^2\bigr)
\notag\\[5pt]
&\quad
- \int \frac{\dd^4 q_1}{(2\pi)^4} \frac{\dd^4 q_2}{(2\pi)^4} \frac{\dd^4 q_3}{(2\pi)^4}\, (2\pi)^4 \delta^4(q_{123})\, \phi(q_1) \phi(q_2) \phi(q_3)\, \frac16 \mu
\notag\\[5pt]
&\quad
- \int \frac{\dd^4 q_1}{(2\pi)^4} \frac{\dd^4 q_2}{(2\pi)^4} \frac{\dd^4 q_3}{(2\pi)^4} \frac{\dd^4 q_4}{(2\pi)^4}\, (2\pi)^4 \delta^4(q_{1234})\, \phi(q_1) \phi(q_2) \phi(q_3) \phi(q_4)\, \frac{1}{24} \lambda \,,
\end{align}
where we have used the abbreviation $q_{1\cdots k} \equiv q_1 + \cdots + q_k$, and $q_1q_2 \equiv q_1\cdot q_2$.

One choice of the metric that satisfies \cref{eqn:Metric} is
\begin{align}
g_{\phi\phi}(p_1, p_2) &= (2\pi)^4 \delta^4(p_{12}) \left( 1 + \frac{m^2}{p_1p_2} \right)
+ \frac13 \mu\, \frac{1}{p_1p_2}\, \phi(-p_{12})
\notag\\[5pt]
&\qquad
+ \frac{1}{12} \lambda\, \frac{1}{p_1p_2} \int \frac{\dd^4 q_1}{(2\pi)^4} \frac{\dd^4 q_2}{(2\pi)^4}\, (2\pi)^4 \delta^4(p_{12} + q_{12})\, \phi(q_1) \phi(q_2) \,.
\label{eqn:MetricExample}
\end{align}
The first and second derivatives of the metric are
\begin{subequations}
\begin{align}
g_{\phi\phi, \phi}(p_1, p_2, p_3) &= \left[ (2\pi)^4 \delta^4(p_{123})\, \frac13 \mu + \frac16 \lambda\, \phi(-p_{123}) \right] \frac{1}{p_1p_2}
\,, \\[5pt]
g_{\phi\phi, \phi\phi}(p_1, p_2, p_3, p_4) &= (2\pi)^4 \delta^4(p_{1234})\, \frac16 \lambda\, \frac{1}{p_1p_2} \,.
\end{align}
\end{subequations}
The inverse metric and the Christoffel connection evaluated at the physical vacuum $\bar\phi=0$ are
\begin{align}
\gbar^{ab} &= (2\pi)^4 \delta^4(p_{ab})\, \frac{p_a^2}{p_a^2-m^2}
\,, \\[5pt]
\overline\Gamma^a_{12} &= -(2\pi)^4 \delta^4(p_a-p_{12})\, \frac{p_a^2}{p_a^2-m^2}\,
\frac16 \mu \left( \frac{1}{p_ap_1} + \frac{1}{p_ap_2} + \frac{1}{p_1p_2} \right) \,.
\end{align}
Here we have started using our abbreviation $\phi^{i_a}(p_a) \to \phi^a$. The Riemann curvature tensor (evaluated at $\bar\phi=0$) can then be computed:
\begin{align}
\Rbar_{1234} &= -\frac12 \left( \gbar_{13,24} - \gbar_{23,14} \right) - \gbar_{ab}\, \Gbar^a_{13}\, \Gbar^b_{24} - (3\leftrightarrow 4)
\notag\\[5pt]
&= (2\pi)^4 \delta^4(p_{1234}) \bigg[ \frac{1}{12} \lambda \left(\frac{1}{p_1p_4} - \frac{1}{p_2p_4} \right)
\notag\\[5pt]
&\hspace{40pt}
-\frac{1}{36} \mu^2\, \frac{p_{13}^2}{p_{13}^2-m^2}
\left( \frac{1}{p_1p_{13}} + \frac{1}{p_3p_{13}} + \frac{1}{p_1p_3} \right) \left( \frac{1}{p_2p_{24}} + \frac{1}{p_4p_{24}} + \frac{1}{p_2p_4} \right)
\bigg] - (p_3 \leftrightarrow p_4) \,.
\label{eqn:RbarExample}
\end{align}

\clearpage
The regular functional derivatives of the action $S$ at $\phi=\bar\phi=0$ are:
\begin{subequations}
\begin{align}
\Sbar_{,12} &= (2\pi)^4 \delta^4(p_{12})\, \bigl(p_1^2 - m^2\bigr)
\,, \\[5pt]
\Sbar_{,123} &= - (2\pi)^4 \delta^4(p_{123})\, \mu
\,, \\[5pt]
\Sbar_{,1234} &= - (2\pi)^4 \delta^4(p_{1234})\, \lambda
\,.
\end{align}
\end{subequations}
The covariant derivatives then follow from the relations
\begin{subequations}
\begin{align}
S_{;12} &= S_{,12} - \G^a_{12} S_{,a}
\,, \\[5pt]
S_{;123} &= S_{,123} - \left( \G^a_{12} S_{,a3} \right)_\text{3\,terms} - \G^a_{123} S_{,a}
\,, \\[5pt]
S_{;1234} &= S_{,1234} -\left( \G^a_{12} S_{,a34}\right)_\text{6\,terms} + \left(\G^a_{12} \G^b_{34} S_{,ab} \right)_\text{3\,terms} - \left( \G^a_{123} S_{,a4} \right)_\text{4\,terms} - \G^a_{1234} S_{,a}
\,.
\end{align}
\end{subequations}
Specifically, we obtain
\begin{subequations}
\begin{align}
\Sbar_{;12} &= \Sbar_{,12} = (2\pi)^4 \delta^4(p_{12})\, (p_1^2 - m^2)
\,,\\[5pt]
\Sbar_{;123} &= 0
\,, \label{eqn:Sbar123Example} \\[5pt]
\Sbar_{;1234} &= \left( p_4^2 + 2p_2p_4 \right) \Rbar_{1234} + \left( 2p_3p_4 \right) \Rbar_{1324}
\,, \label{eqn:Sbar1234Example}
\end{align}
\end{subequations}
where the curvature is given in \cref{eqn:RbarExample}. One can check that these explicit results satisfy the general relations in \cref{eqn:SandR}.

Now let us move on to the amplitudes. The usual Feynman propagator and vertices are (recall $\prop^{ab} S_{,bc} = \delta_c^a$ and $V_{,1\cdots k} = S_{,1\cdots k}$)
\begin{subequations}\label{eqn:DeltaVExample}
\begin{align}
\propbar^{ab} &= (2\pi)^4 \delta^4(p_{ab})\, \frac{1}{p_a^2-m^2}
\,,\\[5pt]
\overline{V}_{123} &= - (2\pi)^4 \delta^4(p_{123})\, \mu
\,,\\[5pt]
\overline{V}_{a12} = \overline{V}_{1a2} = \overline{V}_{12a} &= - (2\pi)^4 \delta^4(p_{a12})\, \mu
\,,\\[5pt]
\overline{V}_{1234} &= - (2\pi)^4 \delta^4(p_{1234})\, \lambda
\,.
\end{align}
\end{subequations}
With these, one can compute the 3-point and 4-point amplitudes from \cref{eqn:MinV} (more specifically \cref{eqn:M3M4inV}):
\begin{subequations}
\begin{align}
(2\pi)^4 \delta^4(p_{123})\, \Amp_{123} = \Mbar_{123} \bigr|_\text{on-shell} &= \overline{V}_{123} \bigr|_\text{on-shell} = - (2\pi)^4 \delta^4(p_{123})\, \mu
\,, \\[8pt]
(2\pi)^4 \delta^4(p_{1234})\, \Amp_{1234} = \Mbar_{1234} \bigr|_\text{on-shell} &= \Big[ \overline{V}_{1234} - \propbar^{ab} \big( \overline{V}_{b41} \overline{V}_{a23} + \overline{V}_{b42} \overline{V}_{1a3} + \overline{V}_{b43} \overline{V}_{12a} \big) \Big] \Bigr|_\text{on-shell}
\notag\\[5pt]
&= - (2\pi)^4 \delta^4(p_{1234}) \left( \lambda + \frac{\mu^2}{p_{12}^2-m^2} + \frac{\mu^2}{p_{13}^2-m^2} + \frac{\mu^2}{p_{14}^2-m^2} \right) \,.
\end{align}
\end{subequations}
On the other hand, our geometrized vertices defined in \cref{eqn:Vcaldef} are
\begin{subequations}
\begin{align}
\Vbar_{123} &= \Sbar_{,123} = - (2\pi)^4 \delta^4(p_{123})\, \mu
\,, \label{eqn:Vcal123Example} \\[5pt]
\Vbar_{a12} \bigr|_\text{on-shell} = \Vbar_{1a2} \bigr|_\text{on-shell} = \Vbar_{12a} \bigr|_\text{on-shell} &= \Big( \Sbar_{,a12} - \Gbar^b_{12} \Sbar_{,ab} \Big) \Bigr|_\text{on-shell} = (2\pi)^4 \delta^4(p_{a12})\, \mu\, \frac{m^2}{3p_1p_2}
\,, \label{eqn:Vcala12Example} \\[5pt]
\Vbar_{1234} \bigr|_\text{on-shell} &= \Big[ \Sbar_{,1234} - \big( \Gbar^a_{12} \Sbar_{,a34} \big)_\text{6\,terms} + \big( \Gbar^a_{12} \Gbar^b_{34} \Sbar_{,ab} \big)_\text{3\,terms} \Big] \Bigr|_\text{on-shell}
\notag\\[5pt]
&= - (2\pi)^4 \delta^4(p_{1234}) \Biggl\{ \lambda + \Biggl[ \frac{\mu^2}{p_{12}^2-m^2} \Biggl[ 1 - \biggl( \frac{m^2}{3p_1p_2} \biggr)^2 \Biggr] \Biggr]_\text{3\,terms} \Biggr\} \,. \label{eqn:Vcal1234Example}
\end{align}
\end{subequations}
We see that they are different from standard vertices in \cref{eqn:DeltaVExample}. However, upon assembling them with the same functions in \cref{eqn:MinV,eqn:MinVcal} (specifically \cref{eqn:M3M4inV} for 3-point and 4-point amplitudes), they do give the same amplitudes:
\begin{subequations}
\begin{align}
\Mbar_{123} \bigr|_\text{on-shell} &= \Vbar_{123} \bigr|_\text{on-shell}
\,, \\[8pt]
\Mbar_{1234} \bigr|_\text{on-shell} &= \Big[ \Vbar_{1234} - \propbar^{ab} \big( \Vbar_{b41} \Vbar_{a23} + \Vbar_{b42} \Vbar_{1a3} + \Vbar_{b43} \Vbar_{12a} \big) \Big] \Bigr|_\text{on-shell} \,.
\end{align}
\end{subequations}
In this example, everything we have demonstrated has been done using on-shell quantities. But we know from the general formalism that the agreement also holds off-shell:
\begin{subequations}
\begin{align}
\Mcal_{123} = V_{123} &= \V_{123}
\,,\\[5pt]
\Mcal_{1234} = V_{1234} - \prop^{ab} \big( V_{b41} V_{a23} + V_{b42} V_{1a3} + V_{b43} V_{12a} \big) &= \V_{1234} - \prop^{ab} \big( \V_{b41} \V_{a23} + \V_{b42} \V_{1a3} + \V_{b43} \V_{12a} \big) 
\,.
\end{align}
\end{subequations}
The geometrization $V\to \V$ is just a reorganization of terms in the amplitudes. The geometrized vertices $\V_{1\cdots k}$ are on-shell covariant under field redefinitions.  This demonstrates the main point of the 
letter.

To complete this example, let us also check the relation between the geometrized vertices $\V_{1\cdots k}$ and the covariant derivatives $S_{;1\cdots k}$. For 3-point vertices with all external indices, we see from \cref{eqn:Vcal123Example,eqn:Sbar123Example} that
\begin{equation}
\Vbar_{123} \bigr|_\text{on-shell} \ne \Sbar_{;123} \bigr|_\text{on-shell} = 0 \,.
\end{equation}
So the 3-point amplitude is not on-shell equivalent to the tensor $S_{;123}$. However, for the geometrized 3-point vertex with one internal index, we see from \cref{eqn:Vcala12Example} that
\begin{equation}
\Vbar_{a12} \bigr|_\text{on-shell} = \Sbar_{;a12} \bigr|_\text{on-shell} = 0
\qquad\text{when}\qquad
m^2=0 \,.
\end{equation}
So the 3-point vertex actually does not enter the geometrized expressions of on-shell higher-point amplitudes in massless theories.

To check the 4-point vertex, we compute $\Sbar_{;1234}|_\text{on-shell}$ from \cref{eqn:Sbar1234Example} and obtain
\begin{align}
\Sbar_{;1234} \bigr|_\text{on-shell} &= - (2\pi)^4 \delta^4(p_{1234})
\Bigg\{ \lambda \bigg[ 1 + \frac{m^2}{6} \left( \frac{1}{p_1p_3} + \frac{1}{p_1p_4} \right) \bigg]
\notag\\[5pt]
&\hspace{40pt}
+ \mu^2 \bigg[ \frac{p_{12}^2-2m^2}{p_{12}^2 (p_{12}^2-m^2)} \left( 1 + \frac{m^2}{3p_1p_2} \right)^2 + \frac{1}{p_{13}^2} \left( 1 + \frac{m^2}{3p_1p_3} \right)^2 + \frac{1}{p_{14}^2} \left( 1 + \frac{m^2}{3p_1p_4} \right)^2 \bigg] \Bigg\} \,.
\end{align}
Comparing this with \cref{eqn:Vcal1234Example}, we see that
\begin{equation}
\Vbar_{;1234} \bigr|_\text{on-shell} = \Sbar_{;1234} \bigr|_\text{on-shell}
\qquad\text{when}\qquad
m^2=0 \,,
\end{equation}
This explicitly demonstrates the relation \cref{eqn:VbarTensor} in the $k=4$ case in our example.

\end{document}